

\input mn
\input epsf


\let\sec=\section
\let\ssec=\subsection


\def\bigstrut{\vrule width0pt height0.35truecm}
\font\japit = cmti10 at 11truept
\def\ss{\scriptscriptstyle\rm}
\def\ref{\parskip =0pt\par\noindent\hangindent\parindent
    \parskip =2ex plus .5ex minus .1ex}
\def\gs{\mathrel{\lower0.6ex\hbox{$\buildrel {\textstyle >}
 \over {\scriptstyle \sim}$}}}
\def\ls{\mathrel{\lower0.6ex\hbox{$\buildrel {\textstyle <}
 \over {\scriptstyle \sim}$}}}
\newcount\equationo
\equationo = 0

\newcount\fred
\fred=0

\def\outeqn#1{\the #1}

\def\leftdisplay#1$${\leftline{$\displaystyle{#1}$
  \global\advance\equationo by1\hfill (\the\equationo )}$$}
\everydisplay{\leftdisplay}

\def\eol{\hfill\break}

\def\kms{\;{\rm km\,s^{-1}}}
\def\kmsmpc{\;{\rm km\,s^{-1}\,Mpc^{-1}}}

\def\mpcoh{\,h^{-1}\,{\rm Mpc}}

\def\japfig#1#2#3{
\beginfigure{#1}
\ifnum #2 = 1 
\epsfxsize=8.4cm
\epsfbox[28 186 488 590]{hizg_fig#1.eps}
\fi
\ifnum #2 = 2 
\epsfxsize=8.4cm
\epsfbox[53 15 465 785]{hizg_fig#1.eps}
\fi
\ifnum #2 = 3 
\epsfxsize=8.2cm
\epsfbox[85 5 460 790]{hizg_fig#1.eps}
\fi
\caption{%
{\bf Figure #1.}
#3
}
\endfigure
}



\def\annrev{ARA\&A}

\def\apj{ApJ}
\def\apjs{ApJS}
\def\mn{MNRAS}

%

\pageoffset{-0.8cm}{0.2cm}




\begintopmatter  

\vglue-2.2truecm
\centerline{\japit Accepted for publication in Monthly Notices of the R.A.S.}
\vglue 1.7truecm

\title{Old high-redshift galaxies and primordial density fluctuation spectra}

\author{J.A. Peacock$^1$, R. Jimenez$^1$, J.S. Dunlop$^2$, I. Waddington$^2$, 
H. Spinrad$^3$, D. Stern$^3$, A. Dey$^4$, R.A. Windhorst$^5$}

\affiliation{$^1$Royal Observatory, \bigstrut Blackford Hill, Edinburgh EH9 3HJ \eol
$^2$Institute for Astronomy, Department of Physics and Astronomy, University 
of Edinburgh, Blackford Hill, Edinburgh EH9 3HJ \eol
$^3$Department of Astronomy, University of California, Berkeley, Ca 94720, USA\eol
$^4$National Optical Astronomy Observatories, 950 North Cherry Avenue, Tucson, Az 85726, USA \eol
$^5$Department of Physics and Astronomy, Arizona State University, Tempe, Az 85287-1504, USA
}

\shortauthor{J.A. Peacock et al.}

\shorttitle{Old high-redshift galaxies and primordial fluctuation spectra}


\abstract{%
We have discovered a population of extremely red 
galaxies at $z\simeq 1.5$ which have apparent
stellar ages of $\gs 3$ Gyr, based on detailed spectroscopy in the
rest-frame ultraviolet. In order for galaxies to have existed
at the high collapse redshifts indicated by these ages, there must be a
minimum level of power in the density fluctuation spectrum
on galaxy scales. This paper compares the required power with
that inferred from other high-redshift populations:
damped Lyman-$\alpha$ absorbers and Lyman-limit
galaxies at $z\simeq 3.2$. If the collapse redshifts for the
old red galaxies are in the range $z_c\simeq 6$ -- 8, there is 
general agreement between the various tracers on the required
inhomogeneity on 1-Mpc scales. This level of small-scale power
requires the Lyman-limit galaxies to be 
approximately $\nu\simeq 3.0$ fluctuations, implying a
very large bias parameter $b\simeq 6$.
If the collapse redshifts of the red galaxies are indeed in the range 
$z_c = 6 - 8$ required for power spectrum consistency, their 
implied ages at $z \simeq 1.5$ are between 3 and 3.8 Gyr for essentially any model 
universe of current age 14 Gyr. The age of these objects as deduced from 
gravitational collapse thus provides independent support for the
ages estimated from their stellar populations.
Such early-forming galaxies are rare, and their contribution to the cosmological
stellar density is consistent with an extrapolation to higher redshifts
of the star-formation rate measured at $z<5$; there is no evidence for a
general era of spheroid formation at extreme redshifts.
}


\maketitle  

\sec{INTRODUCTION}

It is widely believed that the sequence of cosmological
structure formation was hierarchical, originating in 
a density power spectrum with increasing fluctuations on small scales.
The large-wavelength portion of this spectrum
is accessible to observation today through studies of
galaxy clustering in the linear and quasilinear regimes.
However, nonlinear evolution has effectively erased
any information on the initial spectrum for wavelengths
below about 1 Mpc. The most sensitive way of measuring
the spectrum on smaller scales is via the abundances of
high-redshift objects; the amplitude of fluctuations
on scales of individual galaxies governs the redshift
at which these objects first undergo gravitational collapse.

The aim of this paper is to apply these arguments about the
small-scale spectrum to a particularly interesting class
of galaxy which we have recently discovered.
It has long been apparent that a significant fraction of the
optical identifications of 1-mJy radio galaxies are
red and inactive (Windhorst, Kron \& Koo 1984;
Kron, Koo \& Windhorst 1985). More recently,
we have obtained the deep absorption-line spectroscopy
needed in order to prove that these colours result
from a well-evolved stellar population. The minimum
age of the stars can be inferred robustly from spectral
breaks, and gives ages of 3.5 Gyr for 53W091 at
$z=1.55$ (Dunlop et al. 1996; Spinrad et al. 1997), and 4.0 Gyr for
53W069 at $z=1.43$ (Dunlop 1998; Dey et al. 1998). Such ages push the
formation era for these galaxies back to extremely high
redshifts, and it is of interest to ask what level of small-scale
power is needed in order to allow this early formation.
However, the dating of stellar populations rests on complex
modelling, and so it is desirable to have an independent
way of checking whether these high collapse redshifts are
correct. We have carried out such a test, using the fact that
the abundances of early-forming galaxies are sensitive to the
amplitude of the small-scale power spectrum.
Requiring a level of small-scale power consistent
with that implied by other high-redshift objects predicts
a collapse redshift for our red galaxies. From this, we can
predict an age -- which can then be compared with
the age results obtained by analyzing stellar populations.

We shall adopt a standard framework for interpreting the abundances
of high-redshift objects in terms of structure-formation models,
as outlined by Efstathiou \& Rees (1988). 
Under the assumption that the growth
of structure proceeds as a gravitational hierarchy with Gaussian primordial
statistics, the abundance of objects of a given mass is
related directly to the rms density fluctuations on that mass
scale. In Section 2, we summarize the necessary elements of
this Press-Schechter theory.
It will be important to achieve a consistent picture
in this analysis between these
observations of high-redshift density fluctuations and the fluctuation
spectrum at the present deduced from galaxy clustering;
Section 3 summarizes our knowledge of the large-scale spectrum.
We then assemble the data on masses and abundance of
high-redshift galaxies in Section 4, summarizing both our own
results and those of other classes of high-redshift objects.
The implied density fluctuation spectrum
is then discussed in Section 5, where we note that these results
require a high level of bias for rare high-redshift galaxies. 
Finally, we return in Section 6 to the
question of stellar ages in our red radio galaxies, in the
light of the collapse redshifts implied by the constraints
on small-scale density fluctuations.

\sec{PRESS-SCHECHTER APPARATUS}

The formalism of Press \& Schechter (1974) gives a way of calculating
the fraction $F_c$ of the mass in the universe which has collapsed into objects
more massive than some limit $M$:
$$
 F_c(>M,z) = 
 1 - {\rm erf}
\,\left[ {\delta_c \over \sqrt{2}\, \sigma(M)}\right].
$$  
Here, $\sigma(M)$ is the rms fractional density contrast
obtained by filtering the linear-theory density field on the 
required scale. In practice, this filtering is usually performed
with a spherical `top hat' filter 
of radius $R$,  with a corresponding mass of $4 \pi \rho_b R^3/3 $,
where $\rho_b$ is the background density. 
The number $\delta_c$
is the linear-theory critical overdensity, which for a `top-hat'
overdensity undergoing spherical collapse is $1.686$ -- virtually
independent of $\Omega$. This form
describes numerical simulations very well (see e.g. 
Ma \& Bertschinger 1994). 
The main assumption is that the density field obeys Gaussian
statistics, which is true in most inflationary models.
Given some estimate of $F_c$, the number $\sigma(R)$
can then be inferred. Note that for rare objects this is a
pleasingly robust process: a large error in $F_c$ will give
only a small error in $\sigma(R)$, because the abundance is
exponentially sensitive to $\sigma$.

Total masses are of course ill-defined both for real astronomical
objects and clumps of particles in simulations; a better
quantity to use is the velocity dispersion.
Virial equilibrium for a halo of mass $M$ and proper radius $r$ demands
a circular orbital velocity of
$$
V_c^2 = {GM \over r}
$$ 
For a spherically collapsed object this velocity  can be converted directly
into a Lagrangian comoving radius 
which contains the mass of the object within the virialization radius
(e.g. White, Efstathiou \& Frenk 1993)
$$
R / \mpcoh= {2^{1/2}[V_c/100 \kms] \over  \Omega_m^{1/2}(1+z_c)^{1/2} f_c^{1/6}}.
$$
($h\equiv H_0/100 \kmsmpc$).
Here,  $z_c$ is the redshift of virialization; $\Omega_m$ is
the {\it present\/} value of the matter density parameter;
$f_c$ is the density contrast at virialization
of the newly-collapsed object relative
to the background, which is adequately approximated by
$$
f_c=178/\Omega_m^{0.6}(z_c),
$$
with only a slight sensitivity to whether $\Lambda$ is non-zero
(Eke, Cole \& Frenk 1996).

For isothermal-sphere haloes, the velocity dispersion is
$$
\sigma_v=V_c/\sqrt{2}.
$$
Given a formation redshift of interest, and a velocity dispersion, there
is then a direct route to the Lagrangian radius from which the
proto-object collapsed.
It is sometimes argued that the observed stellar velocity dispersion should be
a little `cooler' than that of the dark-matter halo hosting the galaxy
(by a factor $\sqrt{3/2}$ for a $r^{-3}$ stellar density profile
in an isothermal sphere, for example). 
However, this assumes that the dark matter totally
dominates the gravity, whereas real galaxies are baryon
dominated in the centre. Any velocity correction is therefore
likely to be small in practice, and we ignore
the effect.

The Press-Schechter collapsed fraction can now be converted to a differential
number density of objects, $n(M)$, using
$$
M\, n(M) = \rho_b\; {dF_c\over dM}.
$$
In practice, however, one is more likely to measure an integrated
number density $N$ of objects which lie above above some mass
threshold, in which case
$$
F_c \sim {M\, N \over \rho_b} = N\, {4\pi R^3\over 3}.
$$
This just says that the collapsed fraction of the mass is the
fraction of the volume contained in the Lagrangian spheres
around each object.
As argued above, even quite large uncertainties in
$F_c$ can have little effect on the implied value of $\sigma(R)$.
This allows us to neglect the uncertain constant of proportionality
in the above relation, which is
$$
F_c = {\epsilon+2\over \epsilon} \;{M\, N \over \rho_b},
$$
for $F_c\propto M^{-\epsilon}$; similarly, the 
uncertainties in estimating the appropriate value of $R$
from the observed circular velocities are often unimportant.
Strictly, the observations are lower limits: we must make at least
sufficient collapsed objects to site the galaxies under study,
but some objects of this mass may give rise to galaxies of
a different type. However, the results are
highly robust to substantial changes in the assumed abundance,
so we shall treat them as measurements.

The number densities require a cosmological model, and
we quote figures assuming $\Omega=1$.
For specific calculations, we scale to other values of $\Omega$ using
$$
n\; D^2\; dr = n_1\; D_1^2\; dr_1,
$$
where $dr$ is increment of comoving distance
and $D$ is angular-diameter distance ($D=R_0S_k(r)/[1+z]$; see the Appendix).
The scaling of $F_c$ with model is different, because the
inferred density of baryonic material depends on the
element of radial distance only:
$$
F_c\; dr = F_c{}_1\; dr_1.
$$

\sec{POWER SPECTRA FROM GALAXY CLUSTERING}

The small-scale $\sigma(R)$ data have to be related to the
present-day observations of large-scale fluctuations in order
to make a consistent picture. The present linear fluctuation spectrum
is known independent of uncertainties about bias for
$R\gs 3 \mpcoh$ (Peacock 1997).
We summarize here the results of these studies.

We use a dimensionless notation for the power spectrum:
$\Delta^2$ is the contribution to the fractional density variance
per unit $\ln k$.  In the convention of Peebles (1980), this is
$$
\Delta^2(k)\equiv{{\rm d}\sigma^2\over {\rm d}\ln k} ={V\over (2\pi)^3}
\, 4\pi \,k^3\, |\delta_k|^2
$$
($V$ being a normalization volume), and the relation to the correlation function is
$$
\xi(r)=\int \Delta^2(k)\; {dk\over k}\; {\sin kr\over kr}.
$$
Similarly, the variance in fractional 
density contrast averaged over spheres of radius $R$ is
$$
\sigma^2(R)=\int \Delta^2(k)\; {dk\over k}\; W_k^2,
$$
where  $W_k=3(\sin y-y\cos y)/y^3$; $y=kR$.

The twin problems with galaxy clustering are (i) the
power-spectrum measurements are nonlinear, rather than
the linear-theory power spectrum required by the 
Press-Schechter method; (ii) the normalization and
even shape of the galaxy spectrum is biased relative
to that of the mass. The first problem can be dealt
with by calibrating the nonlinear effects using
$N$-body simulations. The second is more difficult,
but soluble in a number of limits. First, the overall
normalization can be determined by the Press-Schechter
method of Section 2, as applied to rich clusters.
This gives a measurement of 
the rms in spheres of radius $8\mpcoh$, on which
there is general agreement: 
$$
\sigma_8 = [0.5 - 0.6]\, \Omega^{-0.56}
$$
(Henry \& Arnaud 1989; White, Efstathiou \& Frenk 1993; 
Viana \& Liddle 1996; 
Eke, Cole \& Frenk 1996). On smaller scales, bias is
expected to steepen the galaxy correlations, but this
effect operates on the {\it nonlinear\/} data, and
so has a small effect on the inferred linear spectrum
for $R\gs 3\mpcoh$ (Peacock 1997).

The resulting spectrum shape appears to be inconsistent with
any variant of pure Cold Dark Matter, and is better
described by Mixed Dark Matter with roughly a 30 per cent
admixture of light neutrinos (e.g. Klypin et al. 1993; 
Peacock 1997; Smith et al. 1997).
We are now interested in seeing how well this spectrum
matches onto the smaller-scale data obtained from
abundances of high-redshift galaxies.

\sec{DATA ON HIGH-REDSHIFT GALAXY ABUNDANCES}

In addition to our red mJy galaxies, two classes of high-redshift
object have been used recently to set constraints on
the small-scale power spectrum at high redshift.

\ssec{Damped Lyman-$\alpha$ systems}

Damped Lyman-$\alpha$ absorbers are systems with HI column densities greater than
$\sim 2\times 10^{24}\; \rm m^{-2}$ (Lanzetta et al. 1991).
If the fraction of baryons in
the virialized dark matter halos equals the global value $\Omega_{\rm B}$,
then data on these systems can be used to infer the total fraction of matter that has 
collapsed into bound structures at high redshifts (Ma \& Bertschinger 1994,
Mo \& Miralda-Escud\'{e} 1994; Kaufmann \& Charlot 1994;
Klypin et al. 1995). The highest
measurement at $\langle z \rangle \simeq 3.2$ implies 
$\Omega_{\rm HI}\simeq 0.0025h^{-1}$
(Lanzetta et al. 1991; Storrie-Lombardi, McMahon \& Irwin 1996).
We take $\Omega_{\ss B}h^2 =0.02$ as a compromise between the lower
Walker et al. (1991) nucleosynthesis estimate and the more recent
estimate of 0.025 from Tytler et al. (1996), giving
$$
F_c = {\Omega_{\rm HI}\over \Omega_{\ss B}} \simeq 0.12h
$$
for these systems.
In this case alone, an explicit value of $h$ is required in order to
obtain the collapsed fraction; we take $h=0.65$.

{

\pretolerance 10000

The photoionizing background prevents virialized gaseous
systems with circular
velocities of less than about $50 \kms$ from cooling efficiently,
so that they cannot contract to the high density contrasts 
characteristic of galaxies (e.g. Efstathiou 1992). We follow
Mo \& Miralda-Escud\'{e} (1994) and use the circular velocity range
50  -- $100\kms$ ($\sigma_v=35$ -- $70\kms$) to model the damped
Lyman alpha systems. Reinforcing the photoionization argument,
detailed hydrodynamic simulations imply that the absorbers
are not expected to be associated with very massive dark-matter haloes 
(Haehnelt, Steinmetz \& Rauch 1997). This assumption is consistent with the
rather low luminosity galaxies detected in association with the absorbers
in a number of cases (Le Brun et al. 1996).

}
\ssec{Lyman-limit galaxies}

Steidel et al. (1996) observed star-forming galaxies between
$z=3$ and 3.5 by looking for objects with a spectral
break redwards of the $U$ band. 
Our treatment of these Lyman-limit galaxies is 
similar to that of Mo \& Fukugita (1996), who compared
the abundances of these objects to predictions from
various models.
Steidel et al. give the comoving density of their galaxies as
$$
N(\Omega=1) \simeq 10^{-2.54} \; (\mpcoh)^{-3}.
$$
This is a high number density, comparable to that of
$L^*$ galaxies in the present Universe. The mass of
$L^*$ galaxies corresponds to collapse of a Lagrangian
region of volume $\sim 1\,\rm Mpc^3$, so the collapsed
fraction would be a few tenths of a per cent if the
Lyman-limit galaxies had these masses.

Direct dynamical determinations of these masses
are still lacking in most cases. Steidel et al. attempt to
infer a velocity width by looking at the equivalent
width of the C and Si absorption lines. These are
saturated lines, and so the equivalent width is
sensitive to the velocity dispersion; values in the
range 
$$
\sigma_v\simeq 180 - 320 \kms
$$ 
are implied. These numbers may measure velocities
which are not due to bound material, in which case
they would give an upper limit to $V_c/\sqrt{2}$ for the
dark halo. A more recent measurement of
the velocity width of the H$\alpha$ emission line in one of these objects
gives a dispersion of closer to $100 \kms$ (Pettini, private
communication), consistent with the median velocity
width for Ly$\alpha$ of $140\kms$ measured in similar
galaxies in the HDF (Lowenthal et al. 1997).
Of course, these figures could underestimate the total velocity
dispersion, since they are dominated by emission from the central regions only.
For the present, we consider the
range of values $\sigma_v = 100$ to $320 \kms$, and
the sensitivity to the assumed velocity will be indicated.
In practice, this uncertainty in the velocity does
not produce an important uncertainty in the conclusions.

\ssec{Red radio galaxies}

We have observed two galaxies at $z=1.43$ and 1.55, 
over an area $1.68\times 10^{-3}\; \rm sr$, so a minimal
comoving density is from one galaxy in this redshift range:
$$
N(\Omega=1) \gs 10^{-5.87} \; (\mpcoh)^{-3}.
$$
This figure is comparable to
the density of the richest Abell clusters, and is thus in reasonable
agreement with the discovery that rich high-redshift
clusters appear to contain radio-quiet examples
of similarly red galaxies (Dickinson 1995).

Since the velocity dispersions of these galaxies are
not observed, they must be inferred indirectly. This
is possible because of the known present-day Faber-Jackson
relation for ellipticals. For 53W091, the large-aperture
absolute magnitude is
$$
M_V(z=1.55\mid \Omega=1) \simeq -21.62 -5 \log_{10} h
$$
(measured direct in the rest frame).
From our Solar-metallicity  models, this would be expected to fade
by about 0.9 mag. between $z=1.55$ and the present, for
an $\Omega=1$ model of present age 14 Gyr
(note that Bender et al. 1996 have observed a shift
in the zero-point of the $M-\sigma_v$ relation out to $z=0.37$
of about the expected size).
If we compare these numbers with the $\sigma_v$ -- $M_V$
relation for Coma ($m-M=34.3$ for $h=1$) taken from
Dressler (1984), this gives velocity dispersions in the range
$$
\sigma_v= 222 \; {\rm to}\; 292 \; \kms.
$$
This is a very reasonable range for a giant elliptical,
and we adopt it hereafter. Assuming low-density models
would increase these figures by an amount smaller
than the above range, so we ignore this additional uncertainty.

We note in passing that
these figures also make a prediction for the metallicity: 
$$
{\rm Mg}_2= 0.32\; {\rm to}\; 0.35,
$$
(Dressler 1984) corresponding to
$$
[{\rm Fe/H}] = 0.11\; {\rm to}\; 0.39,
$$
or a metallicity of between 1.3 and 2.5 times Solar.
Care is needed here, however, because this figure
refers to the {\it nuclear\/} metallicity,
whereas our spectra are effectively total. Given the
metallicity gradients in low-redshift ellipticals,
such a slightly super-Solar nuclear metallicity would
result in an integrated mean metallicity of Solar
at best (e.g. Gonz\'alez \& Gorgas 1996; Buzzoni 1996). 
This means that the use of Solar-metallicity
models in estimating the age of the stellar populations
in these galaxies is consistent.

Having established an abundance and an equivalent circular velocity
for these galaxies, our treatment of them will differ in one
critical way from the Lyman-$\alpha$ and Lyman-limit galaxies.
For these, we take the normal Press-Schechter approach, in which
the systems under study are assumed to be newly born. For
the Lyman-$\alpha$ and Lyman-limit galaxies, this may not be a bad approximation,
since they are evolving rapidly and/or display high levels of star-formation
activity. For the radio galaxies, conversely, this would be a very
poor assumption, since the evidence is that they existed as
discrete systems at redshifts much higher than the $z\simeq 1.5$
where we see them today. Our strategy will therefore be to
apply the Press-Schechter machinery at some unknown formation
redshift, and see what range of redshift gives a consistent
degree of inhomogeneity.

\sec{THE SMALL-SCALE FLUCTUATION SPECTRUM}

\ssec{The empirical spectrum}

Fig. 1 shows the $\sigma(R)$ data which result
from the Press-Schechter analysis, for three
cosmologies. The $\sigma(R)$ numbers measured at various
high redshifts have been translated to $z=0$ using the
appropriate linear growth law for density perturbations
(see Appendix).

\japfig{1}{3}
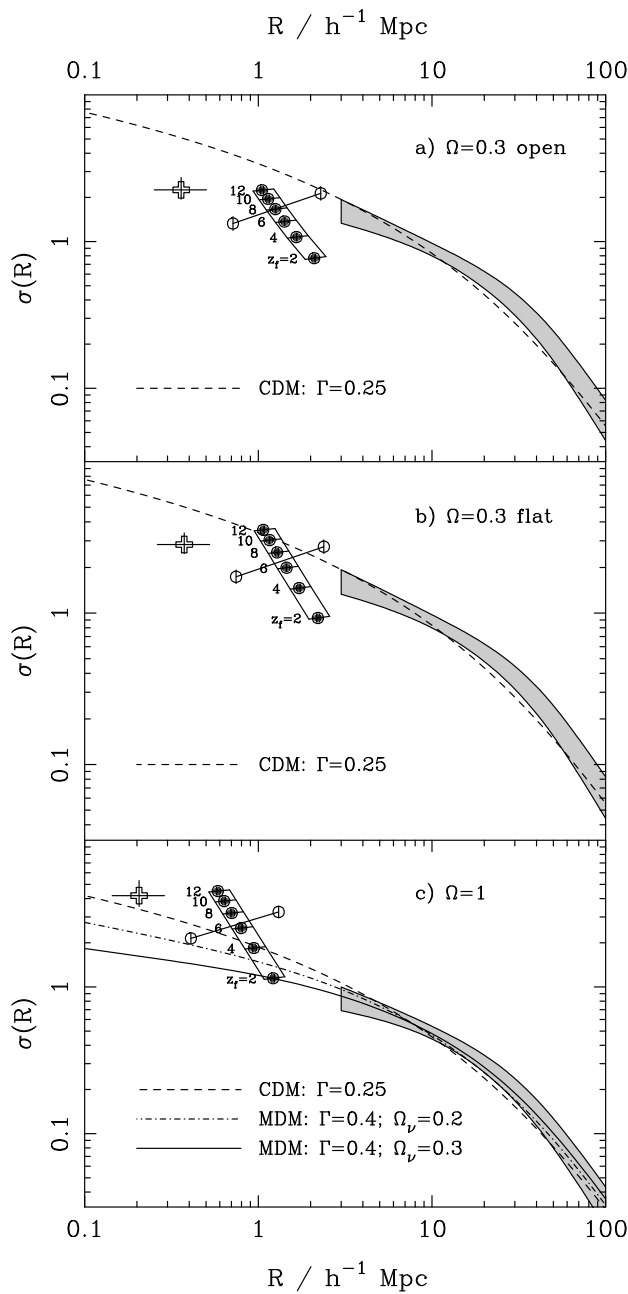
{The present-day linear fluctuation spectrum required
in various cosmologies. This is expressed as $\sigma(R)$:
the fractional rms fluctuation in density averaged in
spheres of radius $R$. The data points are
Lyman-$\alpha$ galaxies (open cross) and
Lyman-limit galaxies (open circles)
The diagonal band with solid points shows red radio
galaxies with assumed collapse redshifts 2, 4, \dots 12.
The vertical
error bars show the effect of a change in abundance by a factor 2.
The horizontal errors correspond to different choices for the
circular velocities of the dark-matter haloes that host the galaxies
($R$ scales linearly with velocity).
The shaded region at large $R$ gives the results inferred
from galaxy clustering.
The solid lines show $\Gamma=0.25$ CDM predictions; for $\Omega=1$
MDM models with $h=0.4$ and $\Omega_\nu=0.2$ and 0.3 (lowest at left)
are also shown.
The large-scale normalization is $\sigma_8=0.55$ for $\Omega=1$
or $\sigma_8=1$ for the low-density models.}

The open symbols give the results for the
Lyman-limit (largest $R$) and Lyman-$\alpha$ (smallest $R$)
systems. The approximately horizontal error bars show the effect of the
quoted range of velocity dispersions for a fixed
abundance; the vertical
errors show the effect of changing the abundance by a factor 2 at fixed
velocity dispersion.
The locus implied by the red radio galaxies sits
in between. The different points show the effects of
varying collapse redshift: $z_c=2, 4, \dots, 12$
[lowest redshift gives lowest $\sigma(R)$]. Clearly, collapse
redshifts of 6 -- 8 are favoured for consistency
with the other data on high-redshift galaxies, independent
of theoretical preconceptions and independent of the
age of these galaxies. This level of power
($\sigma[R]\simeq 2$ for $R\simeq 1 \mpcoh$) is also
in very close agreement with the level of power required to
produce the observed structure in the Lyman alpha forest
(Croft et al. 1997), so there is a good case to be made that
the fluctuation spectrum has now been measured in a 
consistent fashion down to $R\simeq 0.5\mpcoh$.

The shaded region at larger $R$ shows the results
deduced from clustering data (Peacock 1997). The
$\pm 1\sigma$ confidence region was obtained by an approximation to
the fractional error in $\Delta^2(k)$ at $k\simeq 1/R$.
It is clear an $\Omega=1$ universe requires the power spectrum
at small scales to be higher than would be expected on the basis of an
extrapolation from the large-scale spectrum. Depending on assumptions
about the scale-dependence of bias, such a `feature'
in the linear spectrum  may also
be required in order to satisfy the small-scale present-day 
nonlinear galaxy clustering (Peacock 1997).
Conversely, for low-density models, the empirical small-scale
spectrum appears to match reasonably smoothly onto the large-scale data.

\ssec{Comparison with CDM \& MDM}

Fig. 1 also compares the empirical data with various physical power
spectra. A CDM model (using the transfer function of Bardeen et al. 1986) with shape parameter
$\Gamma=\Omega h=0.25$ is shown as a reference for all models.
This has approximately the correct level of small-scale
power, but significantly  over-predicts intermediate-scale clustering, as
discussed in Peacock (1997). The empirical shape is better
described by MDM with $\Omega h\simeq 0.4$ and $\Omega_\nu\simeq 0.3$.
This is the lowest curve in Fig. 1c, reproduced from the
fitting formula of Pogosyan \& Starobinsky (1995; see also
Ma 1996). However, this curve fails to
supply the required small-scale power,
by about a factor 3 in $\sigma$; lowering
$\Omega_\nu$ to 0.2 still leaves a very large discrepancy.
This conclusion is in agreement with e.g. Mo \& Miralda-Escud\'e (1994),
Ma \& Bertschinger (1994), but conflicts slightly with
Klypin et al. (1995), who claimed that the $\Omega_\nu=0.2$
model was acceptable.
This difference arises partly because Klypin et al. adopt a
lower value for $\delta_c$ (1.33 as against 1.686 here), and
also because they adopt the high normalization of
$\sigma_8=0.7$; the net effect of these changes is to boost
the model relative to the small-scale data by a factor of 1.6,
which would allow marginal consistency for the $\Omega_\nu=0.2$
model. MDM models do allow a higher normalization than
the conventional figure of $\sigma_8=0.55$, partly because of
the very flat small-scale spectrum, and also because of the
effects of random neutrino velocities. However, such shifts
are at the 10 per cent level (Borgani et al. 1997a, 1997b), 
and $\sigma_8=0.7$ would probably
still give a cluster abundance in excess of observation.
The consensus of more recent modelling is that even 
$\Omega_\nu=0.2$ MDM is deficient in small-scale power
(Ma et al. 1997; Gardner et al. 1997).

All the models in Fig. 1 assume $n=1$; in fact, consistency with the COBE
results for this choice of $\sigma_8$ requires a significant
tilt for flat models, $n\simeq 0.8$ -- 0.9. 
Over the range of scales probed by large-scale structure,
changes in $n$ are largely degenerate with changes in $\Omega h$,
but the small-scale power is more sensitive to tilt
than to $\Omega h$. Tilting the $\Omega=1$ models is
not attractive, since it increases the tendency for model
predictions to lie below the data. However,
a tilted low-$\Omega$ flat CDM model would agree moderately
well with the data on all scales, with the exception of the
`bump' around $R\simeq 30 \mpcoh$. Testing the reality of this
feature will therefore be an important task for future
generations of redshift survey.

\japfig{2}{1}
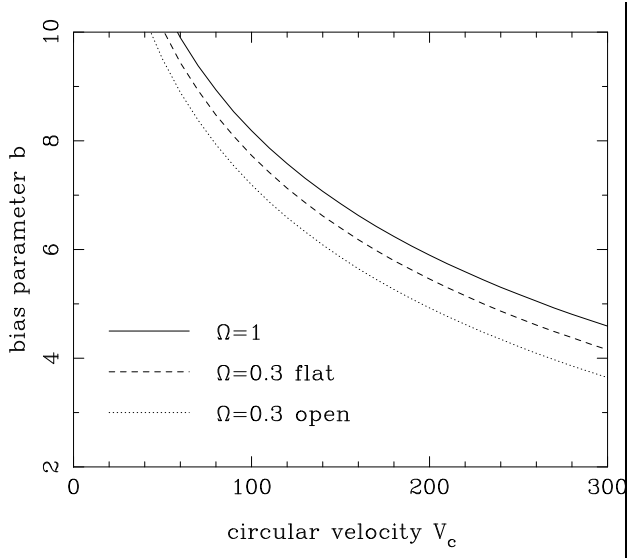
{The bias parameter at $z=3.2$ predicted for the
Lyman-limit galaxies, as a function of their assumed 
circular velocity. Dotted line shows $\Omega=0.3$ open;
dashed line is $\Omega=0.3$ flat; solid line is $\Omega=1$.
A substantial bias in the region of $b\simeq 6$ is predicted
rather robustly.}

\ssec{Limits on high-redshift clustering}

An interesting aspect of these results is that the
level of power on 1-Mpc scales is only moderate:
$\sigma(1\mpcoh)\simeq 2$. At $z\simeq 3$, the
corresponding figure would have been much lower,
making systems like the Lyman-limit galaxies rather
rare. For Gaussian fluctuations, as assumed in the
Press-Schechter analysis, such systems will be
expected to display spatial correlations which are
strongly biased with respect to the underlying mass.
The linear bias parameter depends on the rareness of
the fluctuation and the rms of the underlying field as
$$
b=1+{\nu^2-1\over \nu\sigma}= 1+ {\nu^2-1\over \delta_c}
$$
(Kaiser 1984; Cole \& Kaiser 1989; Mo \& White 1996),
where $\nu = \delta_c/\sigma$, and $\sigma^2$ is the
fractional mass variance at the redshift of interest.

In this analysis, $\delta_c=1.686$ is assumed. 
Variations in this number of order 10 per cent have
been suggested by authors who have studied the
fit of the Press-Schechter model to numerical data.
These changes would merely scale $b-1$ by a small amount;
the key parameter is $\nu$, which is set entirely by
the collapsed fraction. For the Lyman-limit galaxies,
typical values of this parameter are $\nu\simeq 3$,
and it is clear that very substantial values of bias
are expected, as illustrated in Figure 2.

This diagram shows how the predicted bias parameter
varies with the assumed circular velocity, for a number density of
galaxies fixed at the level observed by Steidel et al. (1996).
The sensitivity to cosmological parameter is only
moderate; at $V_c=200\kms$, we have $b\simeq 4.6$, 5.5, 5.8
for the open, flat and critical models respectively.
These numbers scale approximately as $V_c^{-0.4}$, and
$b$ is within 20 per cent of 6 for most plausible
parameter combinations.
Strictly, the bias values determined here are upper
limits, since the numbers of collapsed haloes of this
circular velocity could in principle greatly exceed the
numbers of observed Lyman-limit galaxies. However, the
undercounting would have to be substantial: increasing the
collapsed fraction by a factor 10 reduces the implied bias
by a factor of about 2. A substantial bias seems
difficult to avoid, as has been pointed out in the context of CDM
models by Baugh, Cole \& Frenk (1997).

We now compare these calculations to the
recent detection by Steidel et al. (1997) of strong
clustering in the population of Lyman-limit galaxies at $z\simeq 3$.
The evidence takes the form of a redshift histogram binned
at $\Delta z=0.04$ resolution over a field $8.7' \times 17.6'$ in extent.
For $\Omega=1$ and $z=3$, this probes the density field using a cell with dimensions
$$
{\rm cell} = 15.4 \times 7.6 \times 15.0 \; [\mpcoh]^3.
$$
Conveniently, this has a volume equivalent to a sphere of radius
$7.5 \mpcoh$, so it is easy to measure the bias directly by reference
to the known value of $\sigma_8$. Since the degree of bias is large,
redshift-space distortions from coherent infall are small;
the cell is also large enough that the distortions of small-scale
random velocities at the few hundred $\kms$ level are also small.
Using the model of equation (11) of Peacock (1997) for the
anisotropic redshift-space power spectrum and integrating over
the exact anisotropic window function, we confirm that the above simple
volume argument should be accurate  to a few per cent for reasonable
power spectra:
$$
\sigma_{\rm cell} \simeq b(z=3) \; \sigma_{7.5}(z=3),
$$
where we define the bias factor at this scale. The results of
Mo \& White (1996) suggest that the scale-dependence of bias should be weak.

In order to estimate $\sigma_{\rm cell}$, we have made simulations of 
synthetic redshift histograms,
using the method of Poisson-sampled
lognormal realizations described by Broadhurst, Taylor \& Peacock (1995).
We use a $\chi^2$ statistic to quantify the nonuniformity of the
redshift histogram, and find that $\sigma_{\rm cell}\simeq 0.9$
is required in order for the field of Steidel et al. (1997) to be typical.
It is then straightforward to obtain the bias parameter since, for a 
present-day correlation function $\xi(r)\propto r^{-1.8}$,
$$
\sigma_{7.5}(z=3)=\sigma_8 \times [8/7.5]^{1.8/2} \times 1/4 \simeq 0.146,
$$
implying
$$
b(z=3\mid\Omega=1)\simeq 0.9/0.146 \simeq 6.2.
$$
Steidel et al. (1997) use a rather different analysis which concentrates
on the highest peak alone, and obtain a minimum bias of 6, with a preferred
value of 8. They use the Eke et al. (1996) value of $\sigma_8=0.52$, which
is on the low side of the published range of estimates. Using $\sigma_8=0.55$
would lower their preferred $b$ to 7.6, which is satisfyingly close to our estimate.
Note that, with both these methods, it is much easier to rule out
a low value of $b$ than a high one; given a single field, it is
possible that a relatively `quiet' region of space has been sampled,
and that much larger spikes remain to be found elsewhere. Henceforth,
we assume that the Steidel et al. (1997) field is typical, since there
is evidence that other fields have a similar 
appearance (Steidel, private communication).

Having arrived at a figure for bias if $\Omega=1$, it is easy to
translate to other models, since $\sigma_{\rm cell}$ is observed,
independent of cosmology. For low $\Omega$ models, the cell volume
will increase by a factor $[D^2 dr]/[D_1^2 dr_1]$; comparing with
present-day fluctuations on this larger scale will tend to increase
the bias. However, for low $\Omega$, two other effects increase the
predicted density fluctuation at $z=3$: the cluster constraint
increases the present-day fluctuation by a factor $\Omega^{-0.56}$, and
the growth between redshift 3 and the present will be less than
a factor of 4. Using the Appendix to calculate these corrections, we get
$$
{ b(z=3 \mid \Omega=0.3) \over b(z=3 \mid \Omega=1) } =
\left\{ {0.42\ ({\rm open}) \atop 0.60\ \rlap{({\rm flat})}
\phantom{({\rm open})}} \right. ,
$$
which suggests an approximate scaling as $b\propto \Omega^{0.72}$ (open)
or $\Omega^{0.42}$ (flat).
Multiplying the $\Omega=1$ figure of 6.2 by these factors gives bias values of
2.6 ($\Omega=0.3$ open) or 3.7 ($\Omega=0.3$ flat).
The significance of this observation is thus
to provide the first convincing proof for the reality of galaxy bias: for
$\Omega\simeq 0.3$, bias is not required in the present universe,
but we now see that $b>1$ is needed at $z=3$ for all reasonable
values of $\Omega$.

Comparing these bias values with Fig. 2, we see that
the observed value of $b$ is quite close to the prediction
in the case of $\Omega=1$
-- suggesting that the simplest interpretation of these
systems as collapsed rare peaks may well be roughly correct.
Indeed, for high circular velocities
there is a danger of exceeding the predictions, and it would
create something of a difficulty for high-density models if
a velocity as high as $V_c\simeq 300 \kms$ were to be established as
typical of the Lyman-limit galaxies.
For low $\Omega$, the `observed' bias is lower than the
predictions, so there is no immediate conflict. For a circular
velocity of $200 \kms$, we would need to say that the collapsed fraction
was underestimated by roughly a factor 10 to close the gap in
the case of an open universe. This change in collapsed fraction
increases the values of $\sigma$ in Fig. 1 by a factor
of about 1.5, increasing the `observed' bias by the same factor.
At the same time, this makes $\nu$ smaller, reducing the predicted bias
by about a factor 2 and producing agreement on a bias factor of between 3 and 4.
Such a change in $F_c$ could come about either by postulating that the
conversion from velocity to $R$ is systematically in error, or
by suggesting that there may be many haloes which are not detected
by the Lyman-limit search technique. It is hard to argue that
either of these possibilities are completely ruled out.
Nevertheless, we have reached the paradoxical conclusion that
large-amplitude clustering in the early universe is more naturally
understood in an $\Omega=1$ model, whereas one might have expected the
opposite conclusion.

\sec{AGES AND COLLAPSE REDSHIFTS}

We now return to the red radio galaxies, and
ask if the collapse redshifts 
inferred above are consistent with the
age data on these objects. First bear in mind that
in a hierarchy some of the stars in a galaxy will inevitably
form before the epoch of collapse.
Indeed, some direct observational evidence for the assembly
of galaxies from sub-galactic clumps may now be
starting to emerge (Pascarelle et al. 1996).
At the time of final collapse,
the typical stellar age will be some fraction $\alpha$ of
the age of the universe at that time:
$$
{\rm age} = t(z_{\rm obs}) - t(z_c) + \alpha t(z_c).
$$
We can rule out $\alpha=1$ (i.e. all stars forming in small subunits  just
after the big bang). For present-day
ellipticals, the tight colour-magnitude relation
only allows an approximate doubling of the mass
through mergers since the termination of star formation
(Bower at al. 1992). This corresponds to $\alpha\simeq 0.3$
(Peacock 1991). A non-zero $\alpha$ just corresponds to
scaling the collapse redshift as
$$
{\rm apparent}\ (1+z_c)\propto (1-\alpha)^{-2/3},
$$
since $t\propto (1+z)^{-3/2}$
at high redshifts for all cosmologies.
For example, a galaxy which collapsed at $z=6$ would have
an apparent age corresponding to a collapse redshift of 7.9 for $\alpha=0.3$.

Converting the ages for the galaxies to an apparent collapse
redshift depends on the cosmological model, but particularly on $H_0$.
We can circumvent some of this uncertainty by fixing the age of the
universe. After all, it is of no interest to ask about formation
redshifts in a model with e.g. $\Omega=1$, $h=0.7$ when the whole
universe then has an age of only 9.5 Gyr. If $\Omega=1$ is to be tenable
then either $h<0.5$ against all the evidence or there must be an error
in the stellar evolution timescale. If the stellar timescales
are wrong by a fixed factor, then these two possibilities
are degenerate. It therefore makes sense to measure galaxy ages
only in units of the age of the universe -- or, equivalently, 
to choose freely an apparent Hubble constant which gives the
universe an age comparable to that inferred for globular clusters.
In this spirit, Fig. 3 gives
apparent ages as a function of effective collapse redshift for
models in which the age of the universe is forced to be 14 Gyr
(e.g. Jimenez et al. 1996).

\japfig{3}{1}
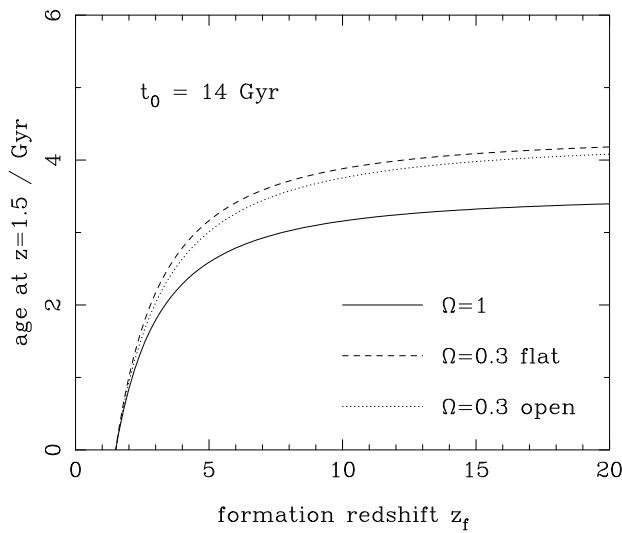
{The age of a galaxy at $z=1.5$, as a function of its
collapse redshift (assuming an instantaneous burst of star formation).
The various lines show $\Omega=1$ [solid]; open $\Omega=0.3$ [dotted];
flat $\Omega=0.3$ [dashed]. In all cases, the present
age of the universe is forced to be 14 Gyr.}

This plot shows that the ages of the red radio galaxies are
not permitted very much freedom. We have argued for
a consistent formation redshift in the range 6 to 8
on abundance grounds, and this clearly
predicts an age of close to 3.0 Gyr for $\Omega=1$,
or 3.7 Gyr for low-density models, irrespective of
whether $\Lambda$ is nonzero.
The age-$z_c$ relation is rather flat, and this gives
a robust estimate of age once we have some idea of $z_c$
through the abundance arguments.
Conversely, it is almost impossible to determine the
collapse redshift reliably from the spectral
data, since a very high precision would be required both
in the age of the galaxy and in the age of the universe.

What conclusions can then be reached about allowed cosmological models?
If we take an apparent $z_c=8$ from the power-spectrum arguments, then
the apparent minimum age of $>4$ Gyr for 53W069 can very nearly be
satisfied in both low-density models
(a current age of 14.5 Gyr would be required), but is unattainable for $\Omega=1$.
In the high-density case, a current age of 17.6 Gyr would be required
to attain the required age for $z_c=8$;
this requires a Hubble constant of $h=0.38$.
As argued above, this conclusion is highly insensitive
to the assumed value of $z_c$.
If the true value of $h$ does turn out to be close to 0.5,
then it might be argued that $\Omega=1$ is consistent with the
data, given realistic uncertainties. The ages for the low-density
models would in this case be large by comparison with the
observed radio-galaxy ages. However, 
the ages obtained by modelling spectra with a single burst
can only be lower limits to the true age for the bulk of the stars;
we could easily be observing an even older burst which is made bluer by
a little recent star formation.
A low $h$ measurement would therefore not rule
out low-density models.

The main conclusion of this paper is thus that the
existence of old radio galaxies at $z=1.5$ 
poses two serious difficulties for an
$\Omega=1$ Universe: (i) a consistent
picture of structure formation through gravitational
instability from Gaussian initial conditions requires
a high formation redshift for these objects, leading to
an old Universe and, particularly, a very small
Hubble constant if the stellar ages of these objects are accepted;
(ii) the shape of the power spectrum is complicated,
with a large change in power between smoothing scales
of $0.5 \mpcoh$ and $5 \mpcoh$; no known model
predicts a spectrum with this shape.
The second difficulty might be avoided through non-Gaussian
statistics, but the first would require our age
estimates for the radio galaxies to be too high
by a factor of about 1.5, which we consider implausible.
The simple solution to these problems is of course
to lower the density parameter, and either open or flat
models with $\Omega\simeq 0.3$ work quite well.
The only counter-argument is that the empirical
fluctuation spectrum then predicts a higher degree
of bias for Lyman-limit galaxies at $z\simeq 3$ than is
observed, whereas prediction and observation match well for
$\Omega=1$. However, this problem disappears if
the $\sigma(R)$ data are increased by a factor $\ls 1.5$;
the question of bias therefore does not significantly affect our
claim that the empirical small-scale fluctuation spectrum is now
measured, once the geometry of the universe is given.

Lastly, it is interesting to note that it has been
possible to construct a consistent picture which
incorporates both the large numbers of star-forming
galaxies at $z\ls 3$ and the existence of old systems
which must have formed at very much larger redshifts.
A recent conclusion from the numbers of Lyman-limit
galaxies and the star-formation rates seen at $z\simeq 1$
has been that the global history of star formation
peaked at $z\simeq 2$ (Madau et al. 1996). This leaves open two possibilities
for the very old systems: either they are the rare precursors
of this process, and form unusually early, or they are
a relic of a second peak in activity at higher redshift,
such as is commonly invoked for the origin of all
spheroidal components.
While we cannot rule out such a bimodal
history of star formation, the rareness of the
red radio galaxies indicates that there is no difficulty
with the former picture. We can demonstrate this quantitatively
by integrating the total amount of star formation at high redshift.
According to Madau et al., The star-formation rate at $z=4$ is
$$
\dot \rho_* \simeq 10^{7.3}h\; M_\odot\,{\rm Gyr}^{-1}\, {\rm Mpc}^{-3},
$$
declining roughly as $(1+z)^{-4}$. This is probably a underestimate
by a factor of at least 3, as indicated by suggestions of dust in
the Lyman-limit galaxies (Pettini et al. 1997), and by the prediction
of Pei \& Fall (1995), based on high-$z$ element abundances.
If we scale by a factor 3, and integrate to find the total density
in stars produced at $z>6$, this yields
$$
\rho_*(z_f>6) \simeq 10^{6.2} M_\odot\,{\rm Mpc}^{-3}.
$$
Since the mJy galaxies have a density of $10^{-5.87}h^3 {\rm Mpc}^{-3}$
and stellar masses of order $10^{11}\, M_\odot$, there is clearly no
conflict with the idea that these galaxies are 
the first stellar systems of $L^*$ size which form
en route to the general era of star and galaxy formation.

\section*{References}

\ref Bardeen J.M., Bond J.R., Kaiser N., Szalay A.S., 1986, {\apj}, {304}, 15 
\ref Baugh C.M., Cole S., Frenk C.S., Lacey C.G., 1997, astro-ph/9703111
\ref Bender R., Ziegler B., Bruzual G., 1996, \apj, 463, L51
\ref Borgani S. Moscardini L., Plionis M., Gorski K.M., Holtzman J., Klypin A., Primak J.R., Smith C.C., Stompor, R., 1997a, New Astronomy, 1, 321
\ref Borgani S., Gardini A., Girardi M., Gottlober S., 1997b, New Astronomy, 2, 119
\ref Bower R.G., Lucey J.R., Ellis R.S., 1992, {MNRAS}, {254}, 601
\ref Broadhurst T.J., Taylor A.N., Peacock J.A., 1995, ApJ, 438, 49
\ref Buzzoni A., 1995, in ``Fresh Views of Elliptical Galaxies'', ASP conf. ser. Vol 86,  eds A. Buzzoni, A. Renzini, A. Serrano, p189
\ref Carroll S.M., Press W.H., Turner E.L., 1992, \annrev, 30, 499
\ref Cole S., Kaiser N., 1989, MNRAS, 237, 1127
\ref Croft R.A.C. et al., 1997, astro-ph/9708018
\ref Dey A., et al., 1998,  for submission to ApJ
\ref Dickinson M., 1995, in ``Fresh Views of Elliptical Galaxies'', ASP conf. ser. Vol 86,  eds A. Buzzoni, A. Renzini, A. Serrano, p283
\ref Dressler A., 1984, \apj, 281, 512
\ref Dunlop J.S., Peacock J.A., Spinrad H., Dey A., Jimenez R., Stern D., Windhorst R.A., 1996, Nat, 381, 581
\ref Dunlop J.S., 1998, astro-ph/9801114
\ref Eke V.R., Cole S., Frenk C.S., 1996, \mn, 282, 263
\ref Efstathiou G., 1992, {\mn}, {256}, 43P
\ref Efstathiou G., Rees M.J., 1988, {\mn}, {230}, 5P
\ref Gardner J.P., Katz N., Weinberg D.H., Hernquist L., 1997, astro-ph/9705118
\ref Gonz\'ales J.J., Gorgas J., 1995, in ``Fresh Views of Elliptical Galaxies'', ASP conf. ser. Vol 86,  eds A. Buzzoni, A. Renzini, A. Serrano, p225
\ref Haehnelt M.G., Steinmetz M., Rauch M., 1997, astro-ph/9706201
\ref Henry J.P., Arnaud K.A., 1991, \apj, 372, 410 
\ref Jimenez R., Thejl P., J\o rgensen U.G., MacDonald J., Pagel B., 1996, \mn, 282, 926
\ref Kaiser N., 1984, ApJ, {284}, L9
\ref Kauffmann G., Charlot S., 1994, ApJ, 430, L97
\ref Klypin A., Holtzman J., Primak J., Reg\H os E., 1993, \apj, 416, 1
\ref Klypin A. et al., 1995, ApJ, 444, 1
\ref Kron R.G., Koo D.C., Windhorst, R.A., 1985, A\&A, 146, 38
\ref Lanzetta K., Wolfe A.M., Turnshek D.A., Lu L., McMahon R.G., Hazard C., 1991, {\apjs}, {77}, 1
\ref Le Brun V., Bergeron J., Boisse P., De Harveng J.M., 1996, A\&A, 321, 733
\ref Lowenthal J.D., et al., 1997, {\apj}, 481, 673
\ref Ma C., Bertschinger E., 1994, {\apj}, {434}, L5
\ref Ma C., 1996, \apj, 471, 13
\ref Ma C., Bertschinger E., Hernquist L., Weinberg D., Katz N., 1997, astro-ph/9705113
\ref Madau P. et al., 1996, MNRAS, 283, 1388
\ref Mo H.J., Miralda-Escud\'{e} J., 1994, {\apj}, {430}, L25
\ref Mo H.J., Fukugita M., 1996, {\apj}, {467}, L9
\ref Mo H.J., White S.D.M., 1996, MNRAS, 282, 1096
\ref Pascarelle S.M., Windhorst R.A., Keel W.C., Odewahn S.C., 1996, Nat, 383, 45
\ref Peacock J.A., 1991, in ``Physical Cosmology'', proc. 2$^{nd}$ Rencontre de Blois, eds A. Blanchard, L. Celnekier, M. Lachi\`eze-Rey \& J. Tr\^an Thanh V\^an (Editions Fronti\`eres), p337
\ref Peacock J.A., 1997, \mn, 284, 885
\ref Peebles P.J.E., 1980, The Large-Scale Structure of the Universe.  Princeton Univ. Press, Princeton, NJ
\ref Pei Y.C., Fall S.M., 1995, ApJ, 454, 69
\ref Pettini M., Steidel C.C., Dickinson M., Kellogg, M., Giavalisco M., Adelberger K.L., 1997, astro-ph/9707200
\ref Pogosyan D.Y., Starobinsky A.A., 1995, \apj, 447, 465
\ref Press W.H., Schechter P., 1974, {\apj}, {187}, 425
\ref Smith C., Klypin A., Gross M., Primack J., Holtzman J., 1997, astro-ph/9702099
\ref Spinrad H., Dey A., Stern D., Dunlop J., Peacock J., Jimenez R., Windhorst R., 1997, ApJ, 484, 581
\ref Steidel C.C., Giavalisco M., Pettini M., Dickinson M., Adelberger K.L., 1996, \apj, 462, L17
\ref Steidel C.C., Adelberger K.L., Dickinson M., Giavalisco M., Pettini M., Kellogg M., 1997, astro-ph/9708125
\ref Storrie-Lombardi L.J., McMahon R.G., Irwin M.J., 1996, MNRAS, 283, L79
\ref Tytler D., Fan X.-M., Burles S., 1996, Nat, 381, 207
\ref Viana P.T., Liddle A.R., 1996, MNRAS 281, 323
\ref Walker T.P., Steigman G., Schramm D.N., Olive K.A., Kang H.S., 1991, {\apj}, {376}, 51
\ref White S.D.M., Efstathiou G., Frenk C.S., 1993, {\mn}, {262}, 1023
\ref Windhorst R.A., Kron R.G., Koo, D.C., 1984, A\&A Suppl., 58, 39

\section*{APPENDIX: FORMULAE FOR GENERAL COSMOLOGIES}

If a nonzero cosmological constant is allowed, not all
of the important cosmological formulae exist as
analytical expressions, but in many cases
accurate approximation formulae may be used:
see Carroll, Press \& Turner (1992).
For convenience, we summarize the necessary expressions here.

In general, it is necessary to distinguish matter ($m$) and vacuum
($v$) contributions to the total density parameter.
Both these parameters and the Hubble parameter
vary with scale factor $a=1/(1+z)$:
$$
H[a]=H_0 \sqrt{\Omega_{\rm v}(1-a^{-2})+\Omega_{\rm m}(a^{-3}-a^{-2})+a^{-2}}.
$$
$$
\Omega_{\rm m}[a]= {\Omega_{\rm m} \over
a + \Omega_{\rm m}(1-a) + \Omega_{\rm v}(a^3-a) },
$$
$$
\Omega_{\rm v}[a]= {a^3 \Omega_{\rm v} \over
a + \Omega_{\rm m}(1-a) + \Omega_{\rm v}(a^3-a) }.
$$

The age of the universe (at a a given epoch, taking the
appropriate redshift-dependent $H$ \& $\Omega$) can
be approximated to a few per cent by
$$
H t={2\over 3}\; |1-f|^{-1/2}\; S_k^{-1} \sqrt{|1-f|\over f},
$$
where $f=0.7\Omega_{\rm m}-0.3\Omega_{\rm v}+0.3$ and $S_k$ is sinh
if $f<1$, otherwise sin.

The increment of comoving distance is
$$
R_0\, dr={[c/H_0]\; dz\over \sqrt{ \Omega_v + \Omega_m(1+z)^3
+ (1-\Omega)(1+z)^2}},
$$
where $\Omega=\Omega_m+\Omega_v$. This integrates to
$$
\eqalign{
R_{0}S_{k}(r) &= {c \over H_{0}} |1-\Omega|^{-1/2} \times \cr
&S_k \left[\int_{0}^{z}\!\!{|1-\Omega|^{1/2} \; dz' \over 
\sqrt{(1-\Omega)(1+z')^{2}+\Omega_v +
\Omega_m(1+z')^{3}}}\right] \cr
}
$$

For the linear growth of density perturbations,
there is a density-dependent suppression of the $\Omega=1$
linear growth law:
$$
\sigma(a)\propto a\,g[\Omega_m(a),\Omega_v(a)],
$$
where a high-accuracy fitting formula is
$$
g(\Omega) ={{5}\over{2}}\Omega_{\rm m}\left[\Omega_{\rm m}^{4/7}-\Omega_{\rm v}+
(1+\Omega_{\rm m}/2)(1+\Omega_{\rm v}/70)\right]^{-1}.
$$
The required growth factor is then
$$
{\sigma(z)\over \sigma(0)} = a\; { g[\Omega_m(a),\Omega_v(a)] \over g[\Omega_m(0),\Omega_v(0)] }.
$$

\bye